\documentclass[sigconf]{acmart}

\usepackage{booktabs}
\usepackage{array}
\usepackage{hyperref}
\usepackage{makecell}
\usepackage{appendix}
\usepackage{float}

\usepackage{ifthen}
\newboolean{showtodos}
\setboolean{showtodos}{true} 
\ifthenelse{\boolean{showtodos}}{
    \usepackage{todonotes}
    \let\xtodo\todo
    \renewcommand{\todo}[1]{\xtodo[inline,color=orange!75]{#1}}
}{
    \newcommand{\todo}[1]{}
}

\usepackage{xcolor}
\usepackage{subcaption}
\definecolor{lightgray}{gray}{0.7}

\AtBeginDocument{%
  }
\setcopyright{acmlicensed}
\copyrightyear{2024}
\acmYear{2024}
\acmDOI{XXXXXXX.XXXXXXX}

\acmConference[KUI '24]{XXI. International Conference on Culture and Computer Science 2024}{October 03--04,
  2024}{Florence, Italy}
\acmISBN{978-1-4503-XXXX-X/18/06}
\begin{document}

\title[MorphoHaptics: Visuohaptic Exploration of Image Datasets]{MorphoHaptics: An Open-Source Tool for Visuohaptic Exploration of Morphological Image Datasets}

\author{Lucas Siqueira Rodrigues}
\email{siqueirl@hu-berlin.de}
\orcid{0000-0001-7675-136X}
\affiliation{%
  \institution{HU Berlin}
  \city{Berlin}
  \country{Germany}
}

\author{Thomas Kosch}
\email{thomas.kosch@hu-berlin.de}
\orcid{0000-0001-6300-9035}
\affiliation{%
  \institution{HU Berlin}
  \city{Berlin}
  \country{Germany}
}

\author{John Nyakatura}
\email{john.nyakatura@hu-berlin.de}
\orcid{0000-0001-8088-8684}
\affiliation{%
  \institution{HU Berlin}
  \city{Berlin}
  \country{Germany}
}

\author{Stefan Zachow}
\email{zachow@zib.de}
\orcid{0000-0001-7964-3049}
\affiliation{%
  \institution{Zuse Institute Berlin}
  \city{Berlin}
  \country{Germany}
}

\author{Johann Habakuk Israel}
\email{israel@htw-berlin.de}
\orcid{0000-0002-8513-6892}
\affiliation{%
  \institution{HTW Berlin}
  \city{Berlin}
  \country{Germany}
}

\renewcommand{\shortauthors}{Siqueira Rodrigues et al.}

\begin{abstract}
Although digital methods have significantly advanced morphology, practitioners are still challenged to understand and process tomographic specimen data. As automated processing of fossil data remains insufficient, morphologists still engage in intensive manual work to prepare digital fossils for research objectives. We present an open-source tool that enables morphologists to explore tomographic data similarly to the physical workflows that traditional fossil preparators experience in the field. We assessed the usability of our prototype for virtual fossil preparation and its accompanying tasks in the digital preparation workflow. Our findings indicate that integrating haptics into the virtual preparation workflow enhances the understanding of the morphology and material properties of working specimens. Our design's visuohaptic sculpting of fossil volumes was deemed straightforward and an improvement over current tomographic data processing methods.
\end{abstract}

\ccsdesc[500]{Human-centered computing~Human computer interaction (HCI)}

\keywords{Haptics, Visuohaptic Integration, Feedback, Data Analysis, Data Exploration, Human-Computer Interaction}


\maketitle

\section{Introduction}
In life sciences, morphology is the study of the form and structure of organisms and their characteristic structural features \cite{thomson_growth_1917}. This scientific discipline has recently undergone profound transformations as a result of the introduction of imaging technologies that revolutionized the study of the structure and function of fossil specimens \cite{pandolfi_editorial_2020}. While practitioners traditionally analyzed fossil structures through the destructive dissection of physical specimens, technological advancements have enabled the non-destructive processing and characterization of fossils, which opened greater possibilities for morphology \cite{elewa_computational_2013}. Computed tomography (CT), magnetic resonance imaging (MRI), and other imaging technologies have enabled the creation of detailed digital models of fossilized specimens, providing extensive data for functional analyses such as finite element analysis (FEA) and computational fluid dynamics (CFD), and allowing researchers to rigorously test hypotheses about the biomechanics and behavior of extinct organisms \cite{cunningham_virtual_2014}. These digital methods have addressed long-standing challenges in morphology, such as technical preparation, namely the complex extraction of fossils from the surrounding matrix \cite{sutton_virtual_2016}. While traditional methods involving mechanical and chemical extraction incur an elevated risk of irreversible damage to delicate fossils, tomographic techniques enable non-destructive digital preparation of specimens as a safer alternative to physical extraction \cite{hoffmann_non-invasive_2014}.

Although the benefits of digital methods in morphology are undeniable, interaction with tomographic fossil data remains "an expensive and time-consuming undertaking" ~\cite{sutton_virtual_2016}. For example, researchers face important issues with the limitations of image segmentation applications when partitioning fossil data to attribute meaning to different regions, as automated segmentation methods are seldom applicable to paleontological datasets~\cite{carvalho_automated_2020}. Indeed, deep learning models require an extensive set of previously analyzed and annotated samples, which is impractical for virtually ungeneralizable fossil datasets \cite{toulkeridou_automated_2023}. Thus, morphology currently relies on semi-automated image segmentation that is performed in data analysis and visualization systems ~\cite{cunningham_virtual_2014}. This procedure involves manual two-dimensional selection of segments of interest, which is performed on numerous individual slices, which is considered by many as a tedious and time-consuming process~\cite{sutton_virtual_2016}. Certain specimen datasets, such as calcite fossils preserved in calcareous sediment, can be particularly difficult to process using thresholding methods due to their low attenuation contrast, increasing the need for manual masking of regions of interest (ROI) \cite{cunningham_virtual_2014}. 

Whereas traditional fossil preparators obtain a wealth of multisensorial cues and use natural gestures while physically extracting specimens from substrates, current digital tools limit interaction with 3D specimens to 2D means. Thus, we present a novel tool that leverages force feedback to essentially emulate physical fossil preparation and improve interaction with fossil image datasets. The visuohaptic exploration of fossil data aims to provide morphologists with a better understanding of specimens' material properties and overall anatomies. We created an algorithm that harnesses voxel intensity values encoded in tomographic data as modulated forces that communicate digital material properties and enhance the formation of accurate multimodal mental representations of fossils. Our contribution introduces an interaction metaphor for the manual processing of digital specimens, which compliments the established additive interaction provided by other tools with more intuitive subtractive methods of volume sculpting to mimic physical extraction in the traditional fossil preparation workflow.

\section{Related Work}
Digital morphology generally leverages a set of tools that perform the functions of importing, visualizing, processing, segmenting, and quantifying image datasets of specimens \cite{maga_digital_2023}. Image processing tools such as ImageJ \cite{abramoff_image_2004} and Fiji \cite{schindelin_fiji_2012} are utilized to process 2D data in preparation for quantitative analyses. Platforms such as MorphoJ offer an integrated environment for various geometric morphometric analyses of 2D and 3D data \cite{klingenberg_morphoj_2011}. Comprehensive general-purpose platforms such as 3D Slicer \cite{pieper_3d_2004} and Amira \cite{stalling_amira_2005} enable practitioners to visualize, segment, and quantify specimens for statistical and comparative analyses. Although such scientific visualization software was not created for the specific needs of morphologists, researchers have extended these tools to streamline their workflows. SlicerMorph fills some of the gaps in the Slicer ecosystem to enable morphologists to retrieve, visualize, measure, and annotate 3D specimens as needed in their workflows \cite{rolfe_slicermorph_2021}. This extension includes modules whose utilities range from the importing of specimen data from MorphoSource \cite{boyer_morphosource_2016} to the automated landmarking of three-dimensional biological structures \cite{porto_alpaca_2021}. 

Although morphologists have created tools that tackle many of their challenges, limited attention has been given to improving digital fossil preparation to approach the intuitiveness of traditional physical extraction. An important effort in this direction is the effort to present tomographic data in an appropriate three-dimensional environment through the implementation of VR in visualization software \cite{besancon_state_2021, korkut_visualization_2023, rodrigues_voxsculpt_2023}. VR has recently been integrated into scientific visualization platforms such as Amira, ParaView, and Slicer \cite{stalling_amira_2005, shetty_immersive_2011, pinter_slicervr_2020}, which is posed to improve interaction with digital specimens through the wealth of spatial and depth cues provided by virtual reality (VR) \cite{bryson_virtual_1996, mccormick_visualization_1988}. Previous research specifically utilized VR to explore morphology \cite{bimber_merging_2002, eckhoff_three-dimensional_2003}, but the interactive capabilities of these solutions are often limited to passive viewing. 

Another important advancement in the direction of natural direct interaction with 3D data has been the integration of haptics into scientific visualization. Haptic Data Visualization has been explored since Iwata and Noma introduced \textit{volume haptization} \cite{iwata_volume_1993} and Avila and Sobierajski presented an interactive method for leveraging force feedback to communicate tomographic volume data \cite{avila_haptic_1996}. The combination of haptics and vision in data exploration has been demonstrated to facilitate the understanding of complex structures \cite{lawrence_synergistic_2004} and to enhance the detection of fuzzier morphological structures, addressing a common issue in volume rendering \cite{palmerius_impact_2009}. Haptic Data Visualization also contributes to better navigation \cite{mendez_haptic-assisted_2005}, ROI selection \cite{wall_quantification_2000}, path following \cite{shen_direct_2019}, and retention \cite{siqueira_rodrigues_comparing_2024}. Open-source platform- and device-agnostic haptic frameworks such as H3D \cite{paneels_prototyping_2013} and Chai3D \cite{f_conti_chai_2003} provide device integration, collision detection, and force calculation algorithms that enable visual and haptic rendering of different types of digital objects and allow researchers to create customized tools that tackle their specific needs. Our research leveraged Chai3D as a haptic framework to fill a gap in harnessing the potential of VR and haptics to tackle specific morphology needs and digital fossil preparation. 

\section{Methodology}
This section describes our design and its technical contributions and details the procedure and artifacts used to evaluate our system. 

\subsection{MorphoHaptics}
We present MorphoHaptics, an interactive system that enables morphologists to explore tomographic image datasets and complete virtual fossil preparation in a manner that aligns with the natural interaction experienced by traditional fossil preparators in their physical fieldwork. Our design introduces an interaction metaphor for the manual processing of digital specimens, complimenting established additive interaction of pixel selection and labeling with more intuitive subtractive methods of volume sculpting. In our prototype, the visuohaptic exploration and processing of fossil specimens is informed by haptics that are modulated by voxel intensity values. Our prototype offers commonly needed functionalities for morphology workflows, such as importing image datasets as tomographic volumes, presenting specimens through volume rendering using color lookup tables, and exporting processed datasets as new volumes and polygonal meshes. Our design also includes a companion module that enables morphologists to examine tomographic data through VR in conjunction with force feedback.


\subsubsection{Voxel-Value Haptics}

We developed an algorithm that enhances Chai3D’s force calculation by integrating averaged voxel intensity values to modulate the haptic feedback intensity. Upon collision, the algorithm samples neighboring voxels and averages their intensity values to compute a scaling factor for the initial force calculation. The process is mathematically represented as:

\begin{equation}
L_{\text{avg}} = \frac{1}{N} \sum_{i=1}^{N} \left( (w_R \cdot R_{\text{norm}} + w_G \cdot G_{\text{norm}} + w_B \cdot B_{\text{norm}}) \cdot A_{\text{norm}} \right)
\end{equation}

where \( N \) is the number of voxels within the spherical region centered at the collision point, and

\begin{equation}
R_{\text{norm}} = \frac{R_i}{255}, \quad G_{\text{norm}} = \frac{G_i}{255}, \quad B_{\text{norm}} = \frac{B_i}{255}, \quad A_{\text{norm}} = \frac{A_i}{255}
\end{equation}

are the normalized red, green, blue, and alpha components of the \( i \)-th voxel, respectively. The weights \( w_R \), \( w_G \), and \( w_B \) are constants representing the contribution of each color channel to the luminosity, with values \( w_R = 0.2126 \), \( w_G = 0.7152 \), and \( w_B = 0.0722 \).

The force \( \mathbf{F} \) applied to the haptic device is then modulated by the average luminosity \( L_{\text{avg}} \):

\begin{equation}
\mathbf{F}' = L_{\text{avg}} \cdot \mathbf{F}
\end{equation}

where \( \mathbf{F} \) is the initial computed force, and \( \mathbf{F}' \) is the adjusted force considering the luminosity.

Our voxel intensity haptics algorithm was designed to inform users of the material properties of digital specimens using information encoded in the data. In tomographic data, voxel values contain radio-density information, which generally correlates to material permeability and hardness. As these material properties are virtually unavailable to vision, translating such information to haptics as the appropriate receiving sense is essential to communicate the characteristics of fossils and to enable morphologists to differentiate visually similar materials. Supplementary haptic cues are poised to enhance the formation of accurate multimodal mental representations of fossils. Our design enables users to toggle voxel-value haptics as they might not find such sensory information to be necessary or optimal for certain tasks or datasets.

\subsubsection{Virtual Dissection through Volume Sculpting}
\begin{figure}[!ht]
\centering
\includegraphics[width=\linewidth]{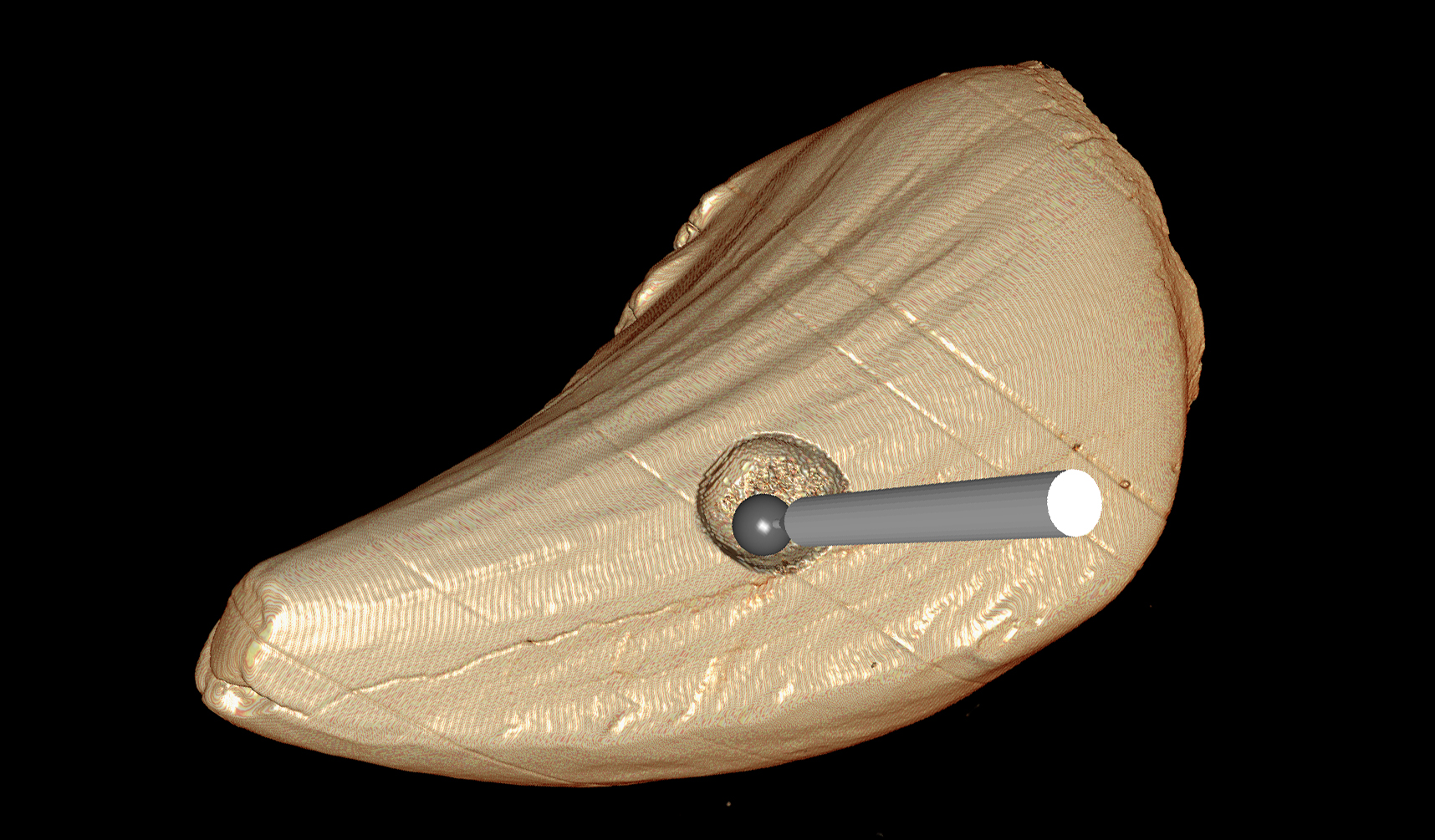}
\caption{Sculpting of an \textit{Ornithischian} tooth. As a user activates the sculpting function and the virtual probe collides with the tomographic volume, affected voxels are removed.} 
\label{volumesculpting}
\end{figure}

Our design utilizes volume sculpting as an interaction metaphor for the manual processing of digital specimens, complimenting established additive interaction of ROI selection  with more intuitive subtractive methods. In manual segmentation, morphologists must refine selections by removing incorrectly assigned voxels whenever thresholding or other selection techniques overstep the boundaries of a region of interest. This process is often done manually on individual tomographic slices, and it is known as the most time-consuming and tedious part of virtual dissection. The issue is particularly pertinent to fossil datasets, as tomographic scans of fossilized structures generally yield low-contrast images, which prevents segmentation algorithms from correctly detecting structures. Our design addresses this issue by enabling three-dimensional volume sculpting that is similar to the matrix removal that technical preparators perform on physical specimens. 

Our tool leverages Chai3D's collision detection system to pinpoint the location of voxels that are touched by the virtual probe during a sculpting gesture, which users can activate while holding the assigned haptic stylus button, as long as the volume sculpting functionality is toggled. During sculpting, targeted voxels are updated to contain zero RGBA values and marked for update during the following rendering iteration. As the 3D texture representing the tomographic volume is updated, graphic and haptic rendering changes are reflected. \autoref{volumesculpting} illustrates the volume sculpting process. In our design, volume sculpting can be assisted by the voxel-value haptics mentioned in the previous subsection. Intensity-modulated haptics enables users to detect changes in material density and prevent unintentional removal of regions of interest. Such complementary sensory information is especially important to volume sculpting because the virtual probe unavoidably occludes its contact area during a collision. As sudden voxel value changes involved in volume sculpting may create undesirable force artifacts, we created an algorithm that smooths haptic rendering by averaging previous and current computed forces. As smoothing forces decreases haptic rendering fidelity, users are able to toggle this function to suit their preferences. 


\subsubsection{User Interface}
\begin{figure}[!ht]
\centering
\includegraphics[width=\linewidth]{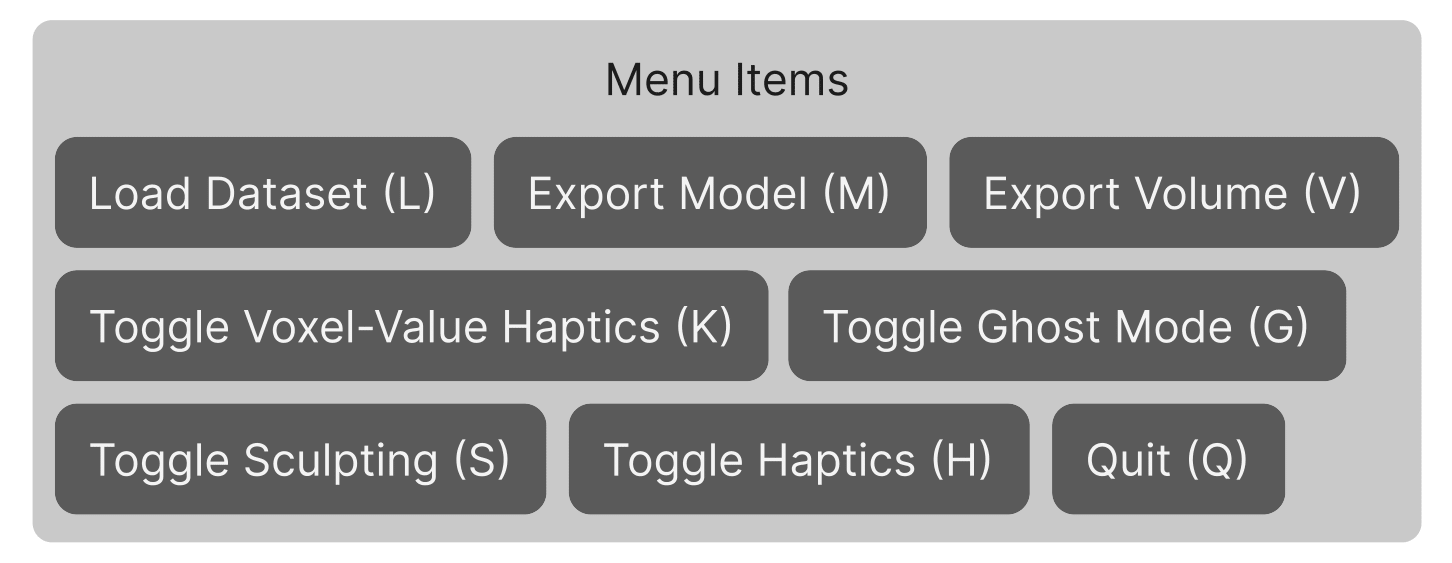}
\caption{MorphoHaptics' UI - Sidebar Menu Items}
\label{userinterface}
\end{figure}

Although our system focuses on direct interaction, we created a basic User Interface (UI) to allow users to control system functionalities. Each functionality has a matching keyboard shortcut, as we foresee a preference for keyboard usage, as users will most likely have their dominant hands occupied with the haptic device, making pointer usage inconvenient. \autoref{userinterface} illustrates the user interface's menu. Menu items that toggle functionalities change their display text (on or off) and color (white or red) to indicate their status. Toggle buttons that are currently in their default mode are displayed in white. In contrast, items that user actions have toggled are shown in red to enable users to understand the system status. Additionally, the UI features a small status bar at the screen’s bottom center to inform users of the system’s status. This space displays temporary messages to indicate the status of transitional actions (e.g., "Exporting Volume...") and static messages to indicate active system status (e.g., "Haptics (off)"). 

\subsubsection{Tomographic Data Handling and Polygonalization}

\begin{figure}[!ht]
\centering
\includegraphics[width=\linewidth]{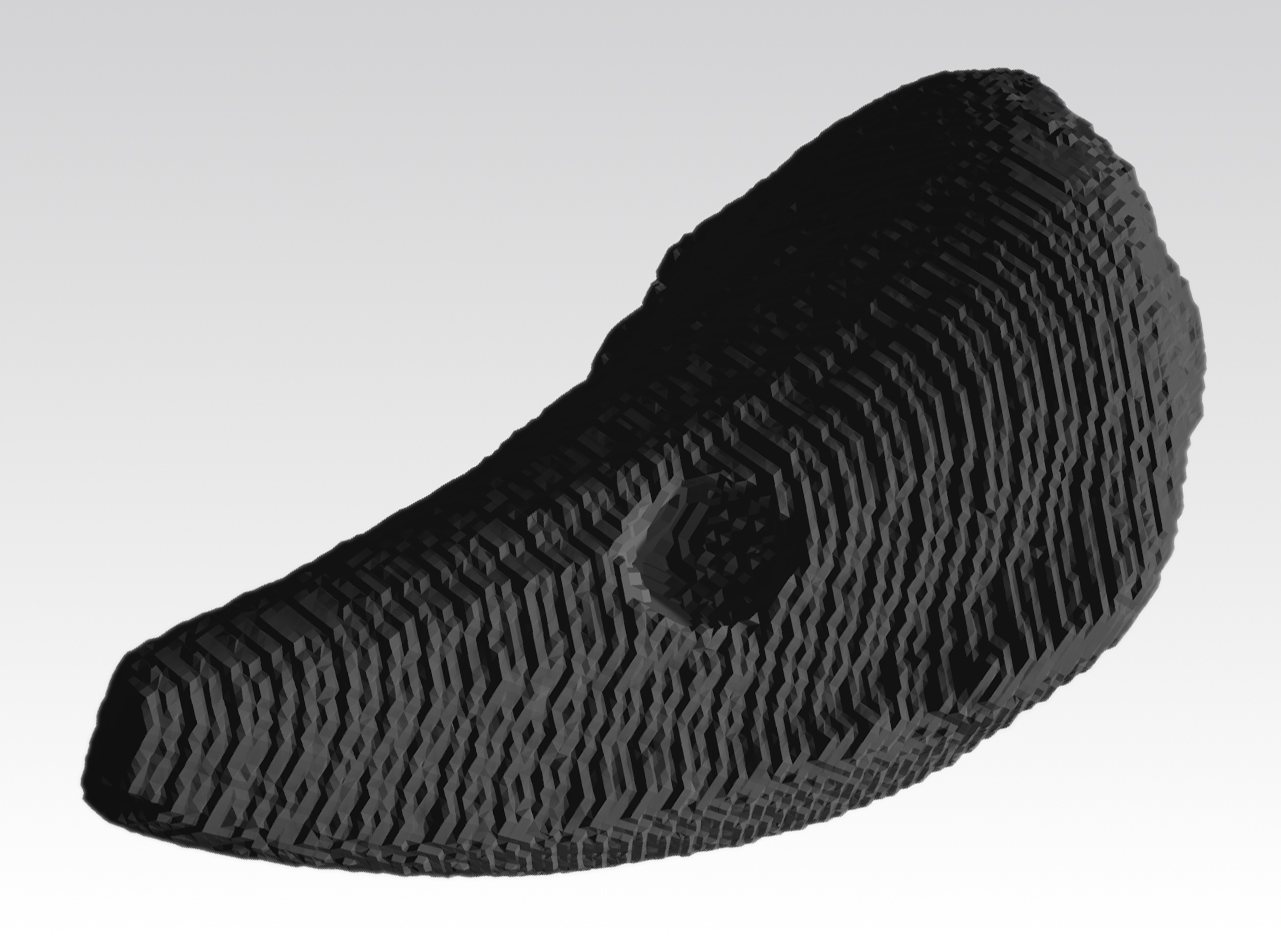}
\caption{Exported model of an \textit{Ornithischian} tooth sculpted in MorphoHaptics. The application polygonized the user-modified tomographic volume and exported a model file.} 
\label{exportedmodel}
\end{figure}

Our application supports the importing and exporting of tomographic volume data and its polygonalization as essential functions in the morphologist toolset. Users can initiate importing through the UI and select the location of a series of image files to be visually and haptically represented in our application as a 3D tomographic volume. Users can later initiate the exporting of modified volumes through a similar process, in which the selected destination folder hosts a sequence of tomographic slices. In addition to tomographic volumes, the application allows users to convert modified volumes into polygonal meshes and export them in a standard model format (STL). \autoref{exportedmodel} shows an example of an exported mesh model.

\subsubsection{Virtual Reality Module}
We created a VR companion module that enables users to perform visuohaptic exploration of fossil datasets in an immersive environment. This module is intended for presenting fossil specimens similarly to what technical preparators experience with real-life physical interaction. Based on the feedback we received on a previous version of this application, we designed this module for sporadic usage, as practitioners are not willing to perform entire workflows in VR nor to spend most of their working hours wearing head-mounted devices. Thus, the VR module does not comprise the functionalities of its Desktop version, being intended as a supplementary companion module.



\subsection{Participants}
\begin{figure}[!ht]
\centering
\includegraphics[width=\linewidth]{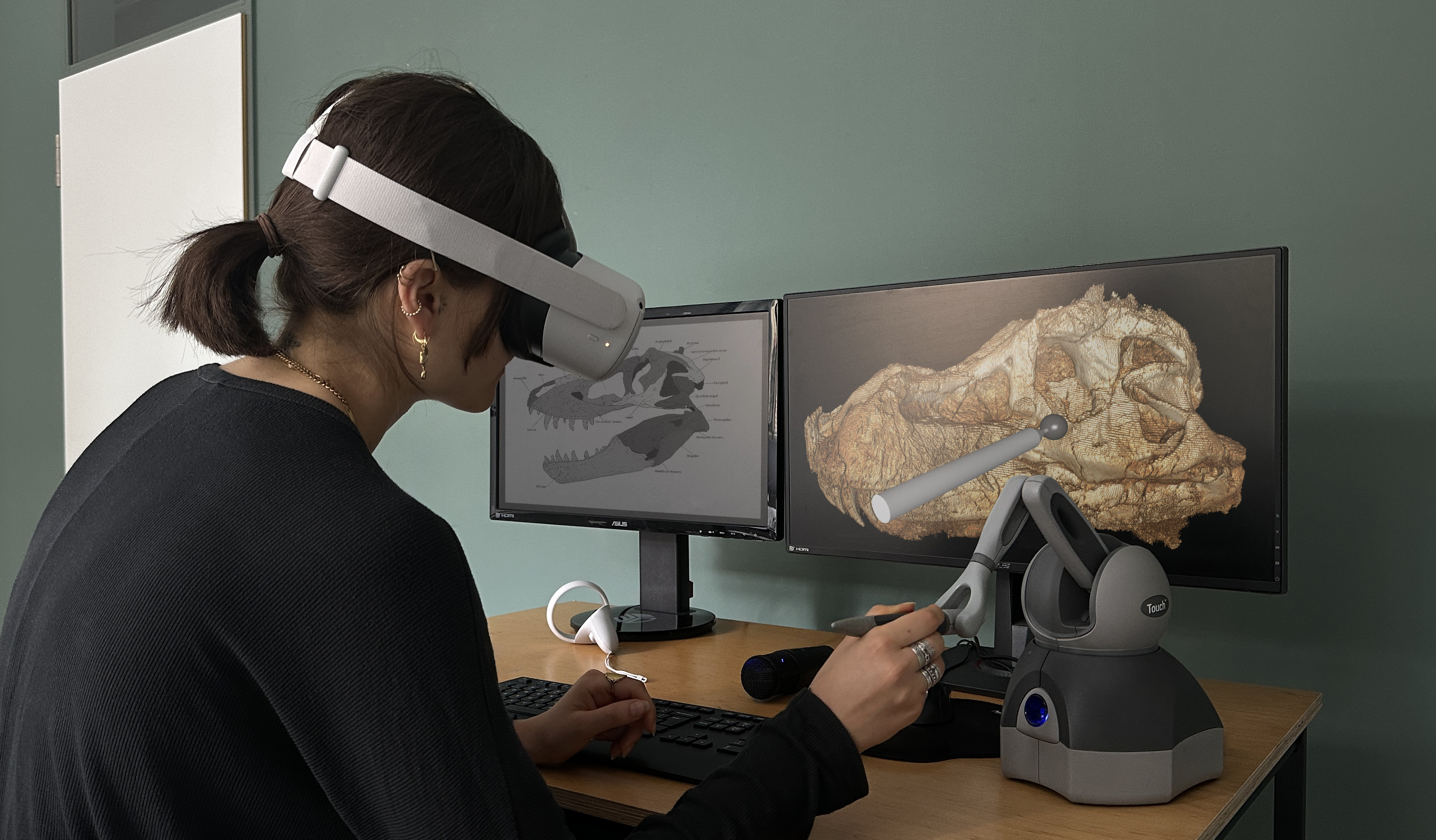}
\caption{A participant exploring a fossil dataset while a diagram is shown on another display (collage).} 
\label{participant}
\end{figure}
Twelve participants completed the study (age: 25.3 $\pm$ 4.03, 9 females and 3 males). Participants were research assistants and interns working in research institutes, as this group comprises the most common profile of novice professionals who manually process tomographic image datasets \cite{siqueira_rodrigues_design_2023}. Participants declared having normal or corrected-to-normal vision and being free of neurological or psychiatric impairments. Eleven participants were right-handed, while one participant was left-handed, to which we accommodated the haptic device's position. None of the participants had professional experience with grounded force-feedback haptic devices. We recruited participants through advertisements posted on online forums and through our extended network of contacts. Participants were informed of our research objectives and were advised to limit socially desirable response bias \cite{nederhof_methods_1985}. Participants provided written informed consent at the beginning of our study, and we monetarily compensated them upon its completion. \autoref{participant} shows a participant interacting with our application. 


\subsection{Evaluation}
To evaluate MorphoHaptics, we utilized specially designed and existing questionnaires that captured users' answers to a series of questions and statements regarding the prototype and its utility to the intended tasks. We created the Visuohaptic Morphology Questionnaire (VHMQ) based on a protocol involving common data exploration and digital fossil preparation tasks. The questionnaire included Likert scale statements, presented on \autoref{VHMQ_graph}, and open-ended questions, detailed on \autoref{tab:questionnaire}. The researchers administered the VHMQ as participants used the application. Upon completion, participants reviewed their answers and confirmed their oral responses. Responses to open-ended questions were recorded in audio and transcribed verbatim using an automated service, followed by eventual manual corrections. After our task protocol, participants completed the ISO 9241-920 Haptic Usability Questionnaire (IHUQ), which we developed based on the aforementioned ISO standard \cite{international_organization_for_standardization_iso_2009}. \autoref{IHUQ_graph} contains the Likert questions, while \autoref{tab:isoquestionnaire} presents IHUQ's open-ended questions. Additionally, we employed the NASA Task Load Index (NASA-TLX) to assess the participant's perceived effort~\cite{10.1145/3582272, 10.1145/3313831.3376766}.

\subsection{Procedure}
\begin{figure*}[!ht]
\centering
\includegraphics[width=\linewidth]{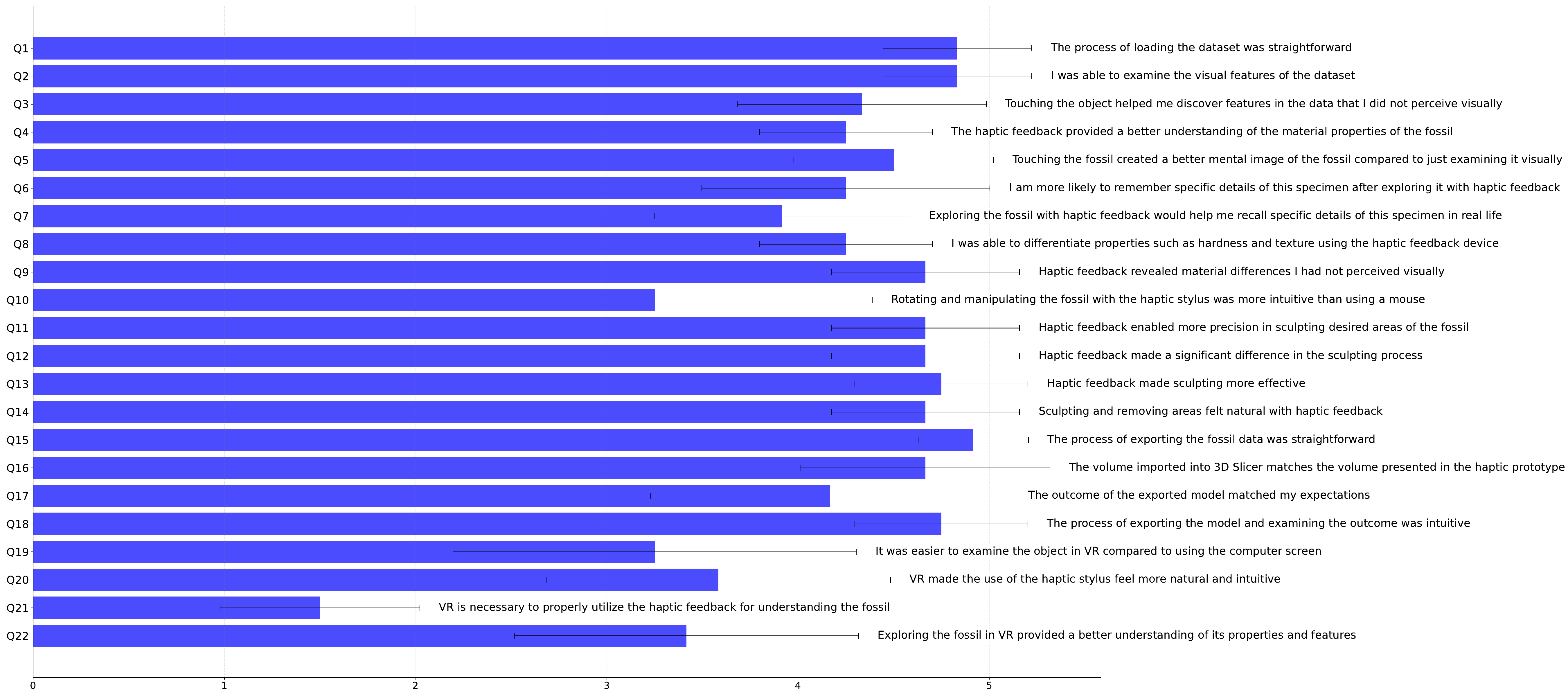}
\caption{Visuohaptic Morphology Questionnaire (VHMQ): Mean (blue) and Standard Deviation (black) for Likert statements.} 
\label{VHMQ_graph}
\end{figure*}

The procedure consisted of nine tasks designed to evaluate aspects of our design, including data loading, visual and visuohaptic exploration, specimen manipulation, volume sculpting, data exporting, and the use of a companion VR application:

\begin{enumerate}
    \item \textbf{Loading the Dataset:} Participants loaded the provided dataset through the System Dialog. Then, they responded to \hyperref[VHMQ_graph]{Q1} about the ease of this process and provided comments (\hyperref[C1]{C1}) on any challenges faced.
    
    \item \textbf{Examining the Loaded Dataset:} Participants visually examined the loaded dataset on the screen, focusing on various fossil parts. We provided participants with a diagram that matched this fossil dataset to guide their visual inspection. They assessed topography and material properties, then responded to \hyperref[VHMQ_graph]{Q2} about the visual examination and provided comments (\hyperref[C2]{C2}) on potential challenges.
    
    \item \textbf{Exploring with Haptic Feedback:} Participants explored the fossil's features using the haptic device's stylus and probed for its morphology and material properties. They answered \hyperref[VHMQ_graph]{Q3} to \hyperref[VHMQ_graph]{Q7}, covering the discovery of features, understanding material properties, mental imagery, memory retention, and recognition of the specimen, and provided comments (\hyperref[C3]{C3} and \hyperref[C4]{C4}) on their experience.
    
    \item \textbf{Differentiating Material Properties:} Participants compared parts of the data using the haptic feedback device to identify properties such as hardness and texture, responded to \hyperref[VHMQ_graph]{Q8} and \hyperref[VHMQ_graph]{Q9} about differentiating these properties, and provided comments (\hyperref[C5]{C5}).
    
    \item \textbf{Rotating and Manipulating the Fossil:} Participants rotated and manipulated the fossil using a mouse and the haptic stylus. Next, they answered \hyperref[VHMQ_graph]{Q10} comparing the intuitiveness of the two methods and provided comments (\hyperref[C6]{C6}).
    
    \item \textbf{Volume Sculpting:} Participants first disabled haptic feedback and sculpted the volume using the stylus, then enabled haptic feedback and repeated the process. They responded to \hyperref[VHMQ_graph]{Q11} to \hyperref[VHMQ_graph]{Q14} about precision, the significance of haptic feedback, effectiveness, and the natural feel of sculpting with haptics, and provided comments (\hyperref[C7]{C7} to \hyperref[C9]{C9}).
    
    \item \textbf{Exporting and Loading Data:} Participants exported the sculpted data to a location, imported it into 3D Slicer, and examined the volume, responding to \hyperref[VHMQ_graph]{Q15} and \hyperref[VHMQ_graph]{Q16} about the exporting process and the match between exported and imported volumes, and providing comments (\hyperref[C10]{C10} and \hyperref[C11]{C11}).
    
    \item \textbf{Exporting a Model and Examining the Outcome:} Participants exported a model and examined the outcome, answering \hyperref[VHMQ_graph]{Q17} and \hyperref[VHMQ_graph]{Q18} regarding whether the outcome matched their expectations and the intuitiveness of the exporting process, and providing comments (\hyperref[C12]{C12} and \hyperref[C13]{C13}).
    
    \item \textbf{Exploring the Dataset in VR:} Participants explored the dataset using the haptic feedback device in VR. They responded to \hyperref[VHMQ_graph]{Q19} to \hyperref[VHMQ_graph]{Q22} about the ease of VR exploration, the natural feel of the haptic stylus, the necessity of VR for utilizing haptics, and the understanding of the fossil's properties in VR, and provided comments (\hyperref[C14]{C14} to \hyperref[C18]{C18}).
\end{enumerate}

Following the completion of Task \#8, which marked the end of the Desktop application evaluation, participants were prompted to fill out the IHUQ and NASA-TLX questionnaires. Next, participants engaged with our companion VR tool and answered questions regarding their experiences. After the VR experience, we requested participants to fill out a separate NASA-TLX form for VR. 

\section{Results}

Here, we report on the quantitative findings of our different evaluation questionnaires. Within the results of each questionnaire, we gather findings into themes regarding the different functionalities of our design. Qualitative statements that elucidate participant feedback are later integrated into themes in the \textit{Discussion}.

\subsection{Visuohaptic Morphology Questionnaire (VHMQ)}

We report on quantitative results for Likert scale statements, whereas answers to open-ended questions are integrated in the following discussion. \autoref{VHMQ_graph} summarizes scores for Likert statements. 

\subsubsection{System Usability and Supporting Functionalities}

Most participants found dataset loading to be straightforward (Q1: $\bar{x} = 4.83$, $\sigma = 0.39$). Participants were generally able to examine the visual features of the fossil dataset (Q2: $\bar{x} = 4.83$, $\sigma = 0.39$). Nevertheless, participants were neutral about the intuitiveness of using the haptic stylus for rotating and manipulating fossils (Q10: $\bar{x} = 3.25$, $\sigma = 1.14$). Participants found fossil volume exporting to be uncomplicated (Q15: $\bar{x} = 4.92$, $\sigma = 0.29$) and its output to be correct (Q16: $\bar{x} = 4.67$, $\sigma = 0.65$). Participants were generally satisfied with exported models' fidelity (Q17: $\bar{x} = 4.17$, $\sigma = 0.94$) while finding the exporting process to be simple (Q18: $\bar{x} = 4.75$, $\sigma = 0.45$).

\subsubsection{Understanding Material Properties and Forming Mental Representations}

Participants generally agreed that touching fossils helped them discover features in the data (Q3: $\bar{x} = 4.33$, $\sigma = 0.65$). Haptic feedback was found to provide a better understanding of the material properties of fossils (Q4: $\bar{x} = 4.25$, $\sigma = 0.45$) and helped participants create better mental representations of specimens (Q5: $\bar{x} = 4.50$, $\sigma = 0.52$). Participants also reported that they were more likely to remember specific details of specimens after exploring them with haptic feedback (Q6: $\bar{x} = 4.25$, $\sigma = 0.75$) while being slightly less certain about whether exploring fossils with haptic feedback would help them recognize and recall specimens' specific details in real life (Q7: $\bar{x} = 3.92$, $\sigma = 0.67$). Participants were able to differentiate properties such as texture and hardness through haptic feedback (Q8: $\bar{x} = 4.25$, $\sigma = 0.45$), which enabled them to discover different material properties that they had not perceived visually (Q9: $\bar{x} = 4.67$, $\sigma = 0.49$).

\subsubsection{Volume Sculpting}

Haptic feedback enabled more precision in sculpting desired fossil areas (Q11: $\bar{x} = 4.67$, $\sigma = 0.49$). Participants reported that haptic feedback made a significant difference in the sculpting process (Q12: $\bar{x} = 4.67$, $\sigma = 0.49$). Furthermore, haptic feedback made sculpting more effective (Q13: $\bar{x} = 4.75$, $\sigma = 0.45$). Sculpting and removing areas felt natural with haptic feedback (Q14: $\bar{x} = 4.67$, $\sigma = 0.49$).

\subsubsection{VR vs. Desktop Environments}

Participants were neutral about their preferences between desktop and VR environments for examining specimens (Q19: $\bar{x} = 3.25$, $\sigma = 1.06$). They were also neutral about whether VR made using the haptic stylus feel more natural and intuitive (Q20: $\bar{x} = 3.58$, $\sigma = 0.90$). Participants generally rejected the necessity of VR for properly utilizing haptic feedback (Q21: $\bar{x} = 1.50$, $\sigma = 0.52$). Lastly, participants were neutral regarding whether VR promoted a better understanding of fossil properties and features (Q22: $\bar{x} = 3.42$, $\sigma = 0.90$).

\begin{figure*}[!ht]
\centering
\includegraphics[width=\linewidth]{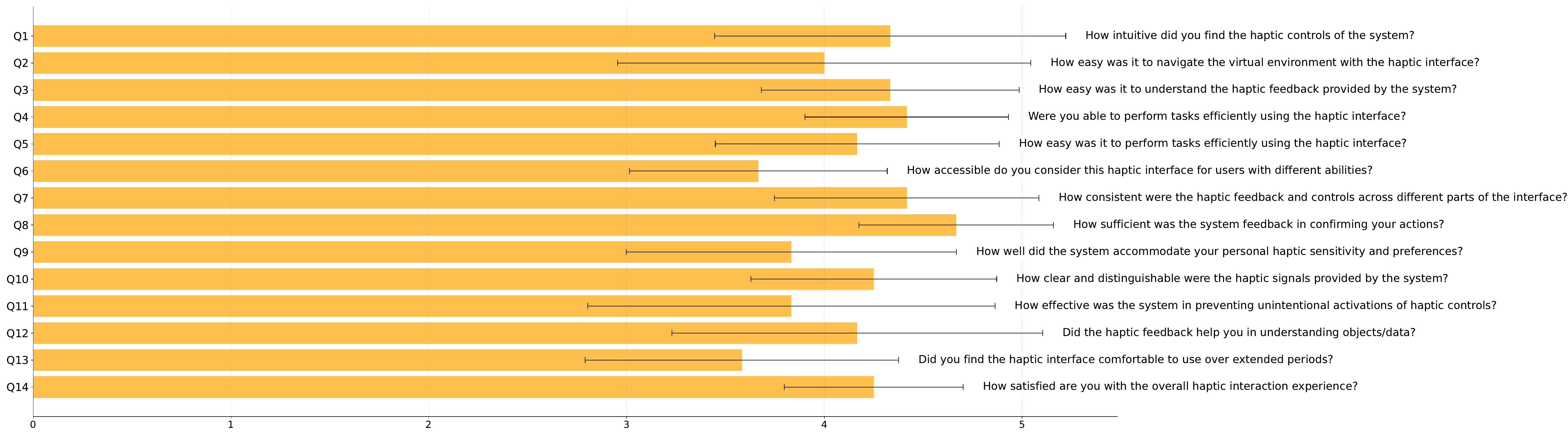}
\caption{ISO 9241-920 Haptic Usability Questionnaire (IHUQ): Mean (orange bars) and Standard Deviation (black lines).} 
\label{IHUQ_graph}
\end{figure*}

\subsection{ISO 9241-920 Haptic Usability Questionnaire (IHUQ)}

Here, we report on quantitative results, while open-ended questions are later reported as qualitative insights in the following discussion. \autoref{IHUQ_graph} summarizes scores for Likert scale questions.

\subsubsection{Usability}

Participants generally found the system's haptic controls to be intuitive (Q1: $\bar{x} = 4.33$, $\sigma = 0.89$). Navigating the virtual environment with the haptic interface was found to be relatively easy (Q2: $\bar{x} = 4.00$, $\sigma = 1.04$). Understanding the system's haptic feedback was also found to be straightforward (Q3: $\bar{x} = 4.33$, $\sigma = 0.65$). Participants could perform tasks efficiently using the haptic interface (Q4: $\bar{x} = 4.42$, $\sigma = 0.51$). The ease of using the haptic interface was rated positively (Q5: $\bar{x} = 4.17$, $\sigma = 0.72$).

\subsubsection{Accessibility}

The accessibility of the haptic interface for users with different abilities received a neutral to positive rate (Q6: $\bar{x} = 3.67$, $\sigma = 0.65$). The consistency of haptic feedback and controls across different interface parts was rated favorably (Q7: $\bar{x} = 4.42$, $\sigma = 0.67$). Participants found the system feedback to be sufficient in confirming their actions (Q8: $\bar{x} = 4.67$, $\sigma = 0.49$). However, the accommodation of personal haptic sensitivity and preferences was rated somewhat lower (Q9: $\bar{x} = 3.83$, $\sigma = 0.83$).

\subsubsection{Feedback and Interaction}

Participants found the system's haptic signals to be clear and distinguishable (Q10: $\bar{x} = 4.25$, $\sigma = 0.62$). The system's effectiveness in preventing unintentional haptic activations was rated moderately (Q11: $\bar{x} = 3.83$, $\sigma = 1.03$). Haptic feedback was found to help understand objects (Q12: $\bar{x} = 4.17$, $\sigma = 0.94$). The comfort of using the haptic interface over extended periods received a neutral rating (Q13: $\bar{x} = 3.58$, $\sigma = 0.79$).

\subsubsection{Overall Experience}

Overall, participants were satisfied with the haptic interaction experience (Q14: $\bar{x} = 4.25$, $\sigma = 0.45$) and would prefer using our haptic interface system over traditional visual systems (Q15: $\bar{x} = 4.67$, $\sigma = 0.49$).

\subsection{NASA-TLX}
\begin{figure}[!ht]
\centering
\includegraphics[width=\linewidth]{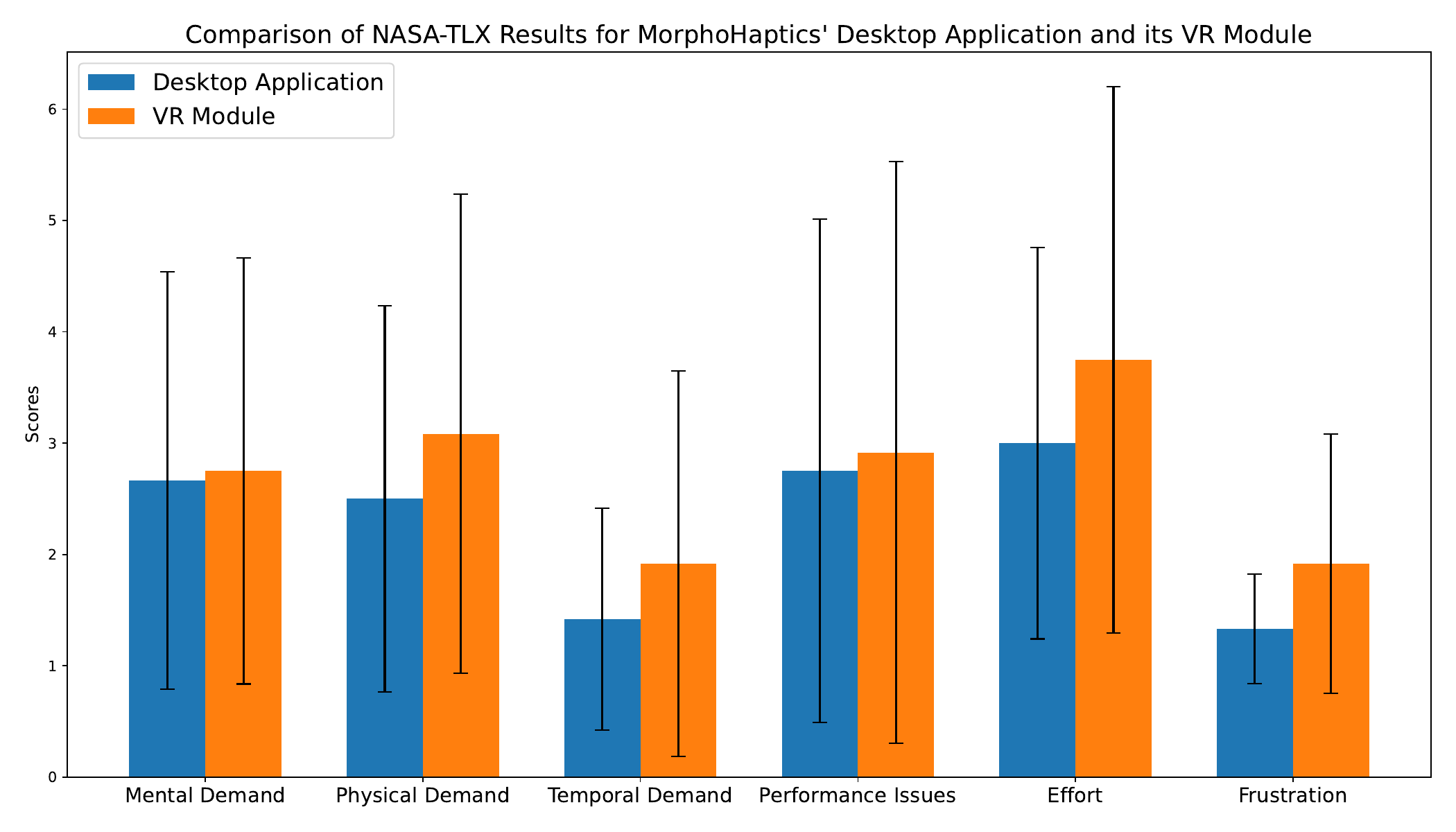}
\caption{Results of the NASA-TLX Workload Assessment Tool for MorphoHaptics and its VR Companion Module.} 
\label{nasa-tlx}
\end{figure}

The NASA-TLX questionnaire results were analyzed to compare the mental workload between MorphoHaptics and its VR Module:

For \textbf{Mental Demand}, both VR and Desktop had a mean of $\bar{x} = 6.5$, with a standard deviation of $\sigma = 3.61$ for each, indicating a similar mental demand perceived in both environments. \textbf{Physical Demand} was slightly higher in VR ($\bar{x} = 2.75$, $\sigma = 1.91$) compared to Desktop ($\bar{x} = 2.67$, $\sigma = 1.87$). This suggests a marginally increased physical effort required in VR. \textbf{Temporal Demand} was also higher in VR ($\bar{x} = 3.08$, $\sigma = 2.15$) compared to Desktop ($\bar{x} = 2.5$, $\sigma = 1.73$), indicating that participants felt more time pressure in the VR environment. \textbf{Performance Issues} were comparable between the two environments, with VR having a mean of $\bar{x} = 2.92$ ($\sigma = 2.61$) and Desktop having a mean of $\bar{x} = 2.75$ ($\sigma = 2.26$). For \textbf{Effort}, VR had a slightly higher mean of $\bar{x} = 3.75$ ($\sigma = 2.45$) compared to Desktop with a mean of $\bar{x} = 3.00$ ($\sigma = 1.76$), indicating that participants felt they exerted more effort in the VR environment. Lastly, \textbf{Frustration} levels were higher in VR ($\bar{x} = 1.92$, $\sigma = 1.16$) compared to Desktop ($\bar{x} = 1.33$, $\sigma = 0.49$), suggesting participants experienced more frustration in the VR application.

\section{Discussion}
Here, we group quantitative results into emerging themes, discuss and interpret the implications of these findings, and integrate answers to open-ended questions as qualitative insights.

\subsection{Understanding Material Properties and Forming Mental Representations}

According to our results, haptic feedback enhanced participants' understanding of fossils' material properties and helped them create better mental representations of specimens. Participants agreed that touching fossils helped them discover features that were not visually perceived. For instance, P1 noted that haptic feedback helped her to "integrate 3D aspects without needing to change views" and to elucidate depth for areas where visual illusions had misled her interpretation of relative profundity to surrounding areas. P3 and P8 noted the added value of texture differentiation for understanding and discriminating between areas of interest. P12 added that haptic cues would be particularly useful in understanding unfamiliar fossil structures. Participants also reported being more likely to remember details of specimens after exploring them with haptic feedback. P8 commented that haptics made it "easier to understand the 3D structure of the fossil", helping him "to remember better where [its] features were". P6 mentioned that haptics provided new depth information for bones she did not know, concluding that touching fossils would "help [her] remember them better". 

\subsection{Voxel-value Haptics}

Our contribution of an algorithm that modulates forces based on radiodensity values reportedly helped participants perceive material properties that are not easily extracted visually. P7 highlighted the value of haptics for discovering material compliance, as she commented on previously having "no clue what the hardness could be like using just sight". P6 stated that such haptic input "helped [her] discover aspects of the object that were not evident from the visual cues". P5 highlighted the utility of material compliance cues in enhancing his understanding of the fossil, as they enabled him to discover that "some [specimen] parts were softer". Our haptic algorithm reportedly improved precision and effectiveness in volume sculpting. P9 stated that volume sculpting particularly benefited from intensity-modulated haptics, stating that when she pressed the stylus "against materials, it was not that easy to separate them, but when [she] started trying to dig in, [she could] see that one material was harder than the other". P11 stated that “in the one fossil [she] carved, [she] could feel a difference in hardness," while simply touching another dataset mostly informed her of its texture. 

\subsection{Volume Sculpting}

Participants indicated that haptics enabled more precision and effectiveness in editing desired volume areas. For example, P8 stated that haptic feedback enabled "precise movements and better control" of his actions while carving a volume, which was similarly reported by P3 and P12. This improvement is particularly important for tasks requiring fine manipulation, demonstrating the practical benefits of the haptic feedback system. Comparing the editing output created with haptics and in its absence, P10 declared being "more satisfied with the outcome with the haptic sculpting". Participants considered haptic sculpting to feel natural, a critical aspect of accepting this novel data interaction metaphor. P5 appreciated the natural aspects of carving while feeling "resistance from the bones" whereas P6 highlighted the importance of feeling "exactly how much pressure [he] was putting in it". P8 considered sculpting to be "a better learning experience than just looking at a 2D image" as commonly done in scientific visualization platforms. 

\subsection{VR vs. Desktop Environments}

Participants expressed contrasting opinions when comparing VR and desktop environments. On average, participants were neutral about their preferred environment and whether VR made haptics feel more natural and intuitive. P2 stated his preference for VR as he could "actually look and do the whole procedure without using something other than the head" to examine specimens, while P11 appreciated the ability to "look at [fossils] from different perspectives [...] and get a better idea of the size of the object". P7 stated that VR promoted "better focus on the task", which is similar to P8's perspective on VR's utility. On the other hand, participants such as P8 "would prefer to use the desktop environment". Others, like P1, would "would use [VR] sporadically" while P6 "would not see [herself] using it" in this context. P7 considered "both VR and Desktop [to be] important", considering the desktop environment for primary use, reckoning that the two environments "complement each other." Participants generally rejected the idea that VR is a prerequisite to benefit from haptic feedback. They were neutral regarding whether exploring fossils in VR provided a better understanding of their properties and features. NASA-TLX results indicated that mental demand and perceived performance were comparable across VR and Desktop environments. However, VR yielded higher physical and temporal demand and greater effort and frustration, although task load scores were low in both environments.

\subsection{Supporting Functionalities}

Participants found the tool easy to use for essential tomographic data functionalities such as loading, examining, and exporting datasets. However, participants P1, P3, and P9 commented that exported polygonal models incurred on loss of fidelity. Although mesh models are expected to feature lower granularity, previewing the export and enabling users to fine-tune detail levels could improve user satisfaction with this feature. Participants differed about the intuitiveness of using the haptic stylus for rotating and manipulating fossils. For example, P9 justified her preference for traditional input by mentioning that she had "been using a computer since [she] was five, so of course it is going to be easier to use a mouse." P4, P10, and P11 made similar comments regarding their mouse proficiency as it contrasted with the challenge of adjusting to a haptic stylus for rotation tasks. On the other hand, P4 and P12 stated that this initial challenge could be overcome with minutes of training. At the same time, P7 found the haptic stylus to excel at combining movements in multiple dimensions, and P10 said using the stylus for manipulation was convenient to avoid switching input devices during active fossil exploration.

\subsection{Usability and Accessibility}

Participants generally found the system's haptic controls and the virtual environment navigation to be intuitive. Understanding the haptic feedback provided by the system was considered to be straightforward. Participants declared being able to perform tasks efficiently using the haptic interface, and statements regarding the ease of performing tasks efficiently using the haptic interface were rated positively. The accessibility of the haptic interface for users with different abilities received a neutral to positive rating, indicating caveats in this assessment. Indeed, P7 noted that our haptic system "is accessible, but it requires training". The consistency of haptic feedback and controls across different interface parts was rated favorably, and participants found the system feedback to be sufficient in confirming their actions. However, the accommodation of personal haptic sensitivity and preferences was rated somewhat lower, which indicates that our system should enable participants to fine-tune force output to match their individual needs for optimal perception. Participants also found the haptic signals provided by the system to be clear, distinguishable, and helpful in understanding objects and data. The system's effectiveness in preventing unintentional activations of haptic controls was rated moderately, even though our UI enabled users to toggle such functions. Moreover, the comfort of using the haptic interface over extended periods received a lower rating, although the workload assessment did not capture this tendency. P8 stated that using the haptic device involved doing "much more work with [his] wrist", which he considered as "physically harder". Nevertheless, participants declared to be satisfied with the haptic interaction experience and would prefer using the haptic interface system over traditional visual systems.

\subsection{Limitations and Future Work}

Participant feedback uncovered several missing auxiliary functionalities that are necessary for the digital morphology workflow. For example, participants inquired about the ability to remap color lookup tables and transfer functions, as a common functionality that biologists expect to have in volume rendering. Another participant noted the missing information about fossil scale relative to the real-world size of displayed specimens. Thus, it is evident that MorphoHaptics has the important limitation of not featuring several functionalities that are common to scientific visualization tools such as Amira or 3D Slicer. As it is not feasible for a novel tool to replicate all the features that such mature platforms contain, a better approach would be to integrate our haptic capabilities into digital environments where morphologists already work. As solutions such as Amira and 3D Slicer allow the community to extend their applications, future work would benefit from incorporating MorphoHaptics' capabilities into established platforms.

\section{Conclusion}

In this study, we presented a novel tool designed to enhance the exploration and understanding of morphological tomographic data through haptic feedback. Our findings indicate that integrating haptics into the morphology workflow significantly improves users’ ability to perceive and differentiate material properties, create accurate mental representations of specimens, and perform precise volume sculpting as digital preparation. Participants evaluated the tool's usability positively, rating it easy to load, examine, and manipulate datasets. Our contribution of an algorithm that modulates forces through voxel intensity values provides a novel approach to communicating the material properties of digital fossils. While VR enhances the overall experience, it is not deemed essential as participants generally found the haptic feedback to be more effective in a desktop setting, which suggests that VR and desktop environments are complementary depending on tasks and user preferences. Despite the promising results, we identified limitations such as the need for finer haptic sensitivity control and the incorporation of additional functionalities commonly found in established scientific visualization tools. Our contribution addresses critical morphology challenges and enhances the exploration and understanding of digital specimens. Moreover, the integration of haptics into a digital morphology tool represents a foundation for future developments in this field.  Future research would benefit from integrating similar haptic capabilities into existing scientific visualization platforms. We make MorphoHaptics available to the community to foster research in this area\footnote{https://github.com/lsrodri/MorphoHaptics}.






\begin{acks}
The author acknowledges the support of the Cluster of Excellence »Matters of Activity. Image Space Material« funded by the Deutsche Forschungsgemeinschaft (DFG, German Research Foundation) under Germany's Excellence Strategy – EXC 2025 – 390648296.
\end{acks}

\bibliographystyle{ACM-Reference-Format}
\bibliography{main}










\clearpage 

\appendix

\begin{table*}[h]
  \section{ISO 9241-920 Haptic Usability Questionnaire (IHUQ): Open-ended Questions}
  \label{tab:isoquestionnaire}
  \begin{tabular}{cp{14cm}}
    \toprule
    \textbf{Code} & \textbf{Question} \\
    \midrule
    C1 & \makecell[l]{\hypertarget{CI1}{} What improvements would you suggest for enhancing the haptic feedback and interaction in this system? \label{CI1}} \\
    C2 & \makecell[l]{\hypertarget{CI2}{} Please describe any specific challenges you faced while using the haptic interface. \label{CI2}} \\
    C3 & \makecell[l]{\hypertarget{CI3}{} How did the haptic feedback contribute to your understanding of objects/data? \label{CI3}} \\
    C4 & \makecell[l]{\hypertarget{CI4}{} Do you have any additional comments or suggestions regarding the haptic interface? \label{CI4}} \\
    \bottomrule
  \end{tabular}
\end{table*}

\begin{table*}
  \section{Visuohaptic Morphology Questionnaire (VHMQ): Open-ended Questions}
  \label{tab:questionnaire}
  \begin{tabular}{cp{14cm}}
    \toprule
    \textbf{Code} & \textbf{Question} \\
    \midrule
    C1 & \makecell[l]{\hypertarget{C1}{} What challenges, if any, did you face while loading the dataset? \label{C1}} \\
    C2 & \makecell[l]{\hypertarget{C2}{} What challenges, if any, did you face while examining the dataset? \label{C2}} \\
    C3 & \makecell[l]{\hypertarget{C3}{} Can you describe any specific features or properties that were easier to understand or remember after \\ \space\space using the haptic feedback? \label{C3}} \\
    C4 & \makecell[l]{\hypertarget{C4}{} How did the haptic feedback influence your overall understanding and memory of the fossil? \label{C4}} \\
    C5 & \makecell[l]{\hypertarget{C5}{} Which properties (e.g., hardness, texture) were most distinguishable with the haptic feedback? \label{C5}} \\
    C6 & \makecell[l]{\hypertarget{C6}{} What did you find more intuitive about using the haptic stylus for rotation and manipulation? \label{C6}} \\
    C7 & \makecell[l]{\hypertarget{C7}{} How did the haptic feedback impact your experience with volume sculpting? \label{C7}} \\
    C8 & \makecell[l]{\hypertarget{C8}{} In both haptic and non-haptic sculpting, how satisfied are you with the outcome of the exported data? \label{C8}} \\
    C9 & \makecell[l]{\hypertarget{C9}{} What differences did you notice between the two methods? \label{C9}} \\
    C10 & \makecell[l]{\hypertarget{C10}{} Did you experience any difficulties in exporting the data? \label{C10}} \\
    C11 & \makecell[l]{\hypertarget{C11}{} Were there any discrepancies between the expected and actual outcomes of the exported data volume? \label{C11}} \\
    C12 & \makecell[l]{\hypertarget{C12}{} Were there any discrepancies between the expected and actual outcomes of the exported model? \label{C12}} \\
    C13 & \makecell[l]{\hypertarget{C13}{} How intuitive did you find the process of exporting and examining the model? \label{C13}} \\
    C14 & \makecell[l]{\hypertarget{C14}{} What aspects of the VR experience did you find most helpful or intuitive? \label{C14}} \\
    C15 & \makecell[l]{\hypertarget{C15}{} Were there any challenges or discomforts you experienced while using VR? \label{C15}} \\
    C16 & \makecell[l]{\hypertarget{C16}{} How did the VR environment impact your ability to understand and interact with the fossil compared \\ \space\space to the desktop environment? \label{C16}} \\
    C17 & \makecell[l]{\hypertarget{C17}{} Did you personally prefer to explore the specimen using the Desktop or VR experience? \label{C17}} \\
    C18 & \makecell[l]{\hypertarget{C18}{} Would you imagine using VR in your current working day? \label{C18}} \\
    \bottomrule
  \end{tabular}
\end{table*}

\begin{table*}
  \section{NASA Task Load Index - Workload Assessment Tool (NASA-TLX)}
  \label{tab:nasa_tlx}
  \begin{tabular}{cp{14cm}}
    \toprule
    \textbf{Code} & \textbf{Question} \\
    \midrule
    TLX1 & \makecell[l]{\hypertarget{TLX1}{} [Mental Demand] How mentally demanding was the task? \label{TLX1}} \\
    TLX2 & \makecell[l]{\hypertarget{TLX2}{} [Physical Demand] How physically demanding was the task? \label{TLX2}} \\
    TLX3 & \makecell[l]{\hypertarget{TLX3}{} [Temporal Demand]How hurried or rushed was the pace of the task? \label{TLX3}} \\
    TLX4 & \makecell[l]{\hypertarget{TLX4}{} [Performance] How successful were you in accomplishing what you were asked to do? \label{TLX4}} \\
    TLX5 & \makecell[l]{\hypertarget{TLX5}{} [Effort] How hard did you have to work to accomplish your level of performance? \label{TLX5}} \\
    TLX6 & \makecell[l]{\hypertarget{TLX6}{} [Frustration] How insecure, discouraged, irritated, stressed, and annoyed were you during the task? \label{TLX6}} \\
    \bottomrule
  \end{tabular}
\end{table*}

\end{document}